\documentclass[aps,pra,twocolumn,showpacs,superscriptaddress]{revtex4}
\usepackage{amsmath,graphicx,color}
\usepackage{times}
\begin{document}
%% $Date: 2010-08-01 17:33:06+09 $
%% $Revision: 4.18 $
%%
\newcommand\ket[1]{\left|\textstyle{#1}\right\rangle}
\newcommand\braket[1]{\left\langle\textstyle{#1}\right\rangle}
\newcommand\half{\frac{1}{2}}
\newcommand\calD{\mathcal{D}}
\newcommand\calE{\mathcal{E}}
\newcommand\Tr{\operatorname{Tr}}
\newcommand\sech{\operatorname{sech}}
\newcommand\cl{\mathrm{cl}}
\newcommand\red{\color{red}}
\newcommand\blue{\color{blue}}

\title{Variational study of a two-level system coupled to a harmonic
  oscillator in a ultra-strong coupling regime}

\author{Myung-Joong Hwang}%
\affiliation{Department of Physics, Pohang University of Science and
  Technology, Pohang 790-784, Korea}%

\author{Mahn-Soo Choi}%
\email{choims@korea.ac.kr}%
\affiliation{Department of Physics, Korea University, Seoul 136-713,
  Korea}%

\begin{abstract}
The non-classical behaviors of a two-level system coupled to a harmonic
oscillator are investigated in the ultra-strong coupling regime. We revisit
the variational solution of the ground state and find that the existing
solutions do not account accurately for non-classical effects such as
squeezing. We suggest a new trial wave function and demonstrate that it has an
excellent accuracy on the quantum correlation effects as well as on energy.
\end{abstract}
\pacs{03.67.Lx, 42.50.Pq, 42.50.Dv, 85.25.Cp}
\maketitle

% \section{Introduction}

A two-level system interacting with a harmonic oscillator appears in
various fields in physics, ranging from an atom coupled to a photon
inside of an optical cavity~\cite{Raimond01a} to a cooper-pair box
coupled to a nanomechanical oscillator~\cite{Lahaye09a}. A
theoretical model that describes the system is expressed by the
Hamiltonian,
\begin{equation}
\label{Paper::eq:1}
H = \omega_0 a^\dag a + \frac{1}{2}\Omega\sigma_z -
\lambda\left(a+a^\dag\right)\sigma_x + \lambda^2
\end{equation}
where $\omega_0$ is the energy of the oscillator, $\Omega$ is the spin level
splitting, and $\lambda$ is the coupling strength. The Pauli matrices
characterize the two level system, while $a$ and $a^\dagger$ denote the boson
operators.

The Hamiltonian (\ref{Paper::eq:1}) reveals completely different physics at
different scales of $\omega_0$, $\Omega$ and $\lambda$. For example, a cavity
quantum electrodynamics (QED) operates in a regime where
$\lambda,|\omega_0-\Omega|\ll\omega_0,\Omega$, under which the
counter-rotating terms of the Hamiltonian (\ref{Paper::eq:1}),
$a^\dagger\sigma^+$ and $a\sigma^-$, can be neglected.  Within this rotating
wave approximation (RWA), the model reduces to a Jaynes-Cummings model and is
solvable exactly~\cite{Jaynes63a}.  The ground state is a simple direct
product of the ground states of the oscillator and the spin.  On the other
hand, in the so-called \emph{ultra-strong coupling} regime
($\Omega\gg\omega_0$ and $\lambda\sim\sqrt{\Omega\omega_0}$), there arise
interesting quantum effects in the ground state; the two-level system and the
oscillator are entangled. The degree of entanglement increases monotonically
as a function of the \emph{dimensionless coupling constant}
$g\equiv2\lambda/\sqrt{\Omega\omega_0}$. It also shows a squeezing
effect~\cite{Yuen76a}, that is, the variance of momentum quadrature in the
ground state becomes smaller than the uncertainty
minimum~\cite{Zazunov06a,Hwang10a,Ashhab10a}.
Hereafter, we will set $\omega_0=1$ for simplicity.

The aforementioned ultra-strong regime has recently attracted much attention
because of the possibility to realize it experimentally in a circuit QED
system~\cite{Devoret07a,Bourassa09a,Ashhab10a,Zueco09a}.  Furthermore, as
rather a theoretical problem, the ultra-strong coupling regime when $\Omega$
becomes $\infty$ has been also studied extensively because it shows a phase
transition-like behavior in the entanglement between the oscillator and the
qubit in the ground
state~\cite{Levine04a,Hines04a,Liberti06a,Meaney10a,Irish07a,1004.2801}.

\begin{figure}
\begin{tabular}{cc}
\includegraphics*[width=42mm]{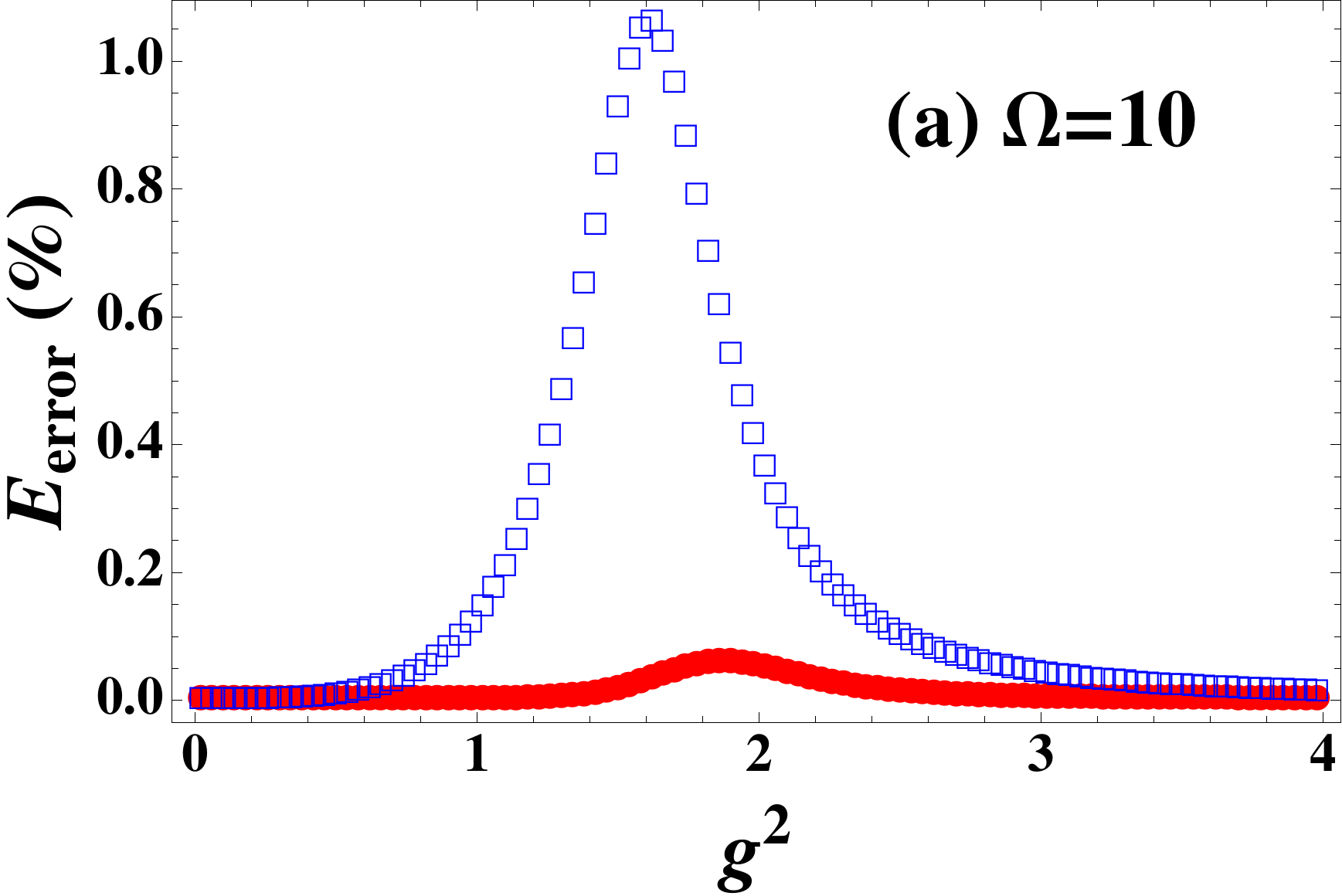} &
\includegraphics*[width=42mm]{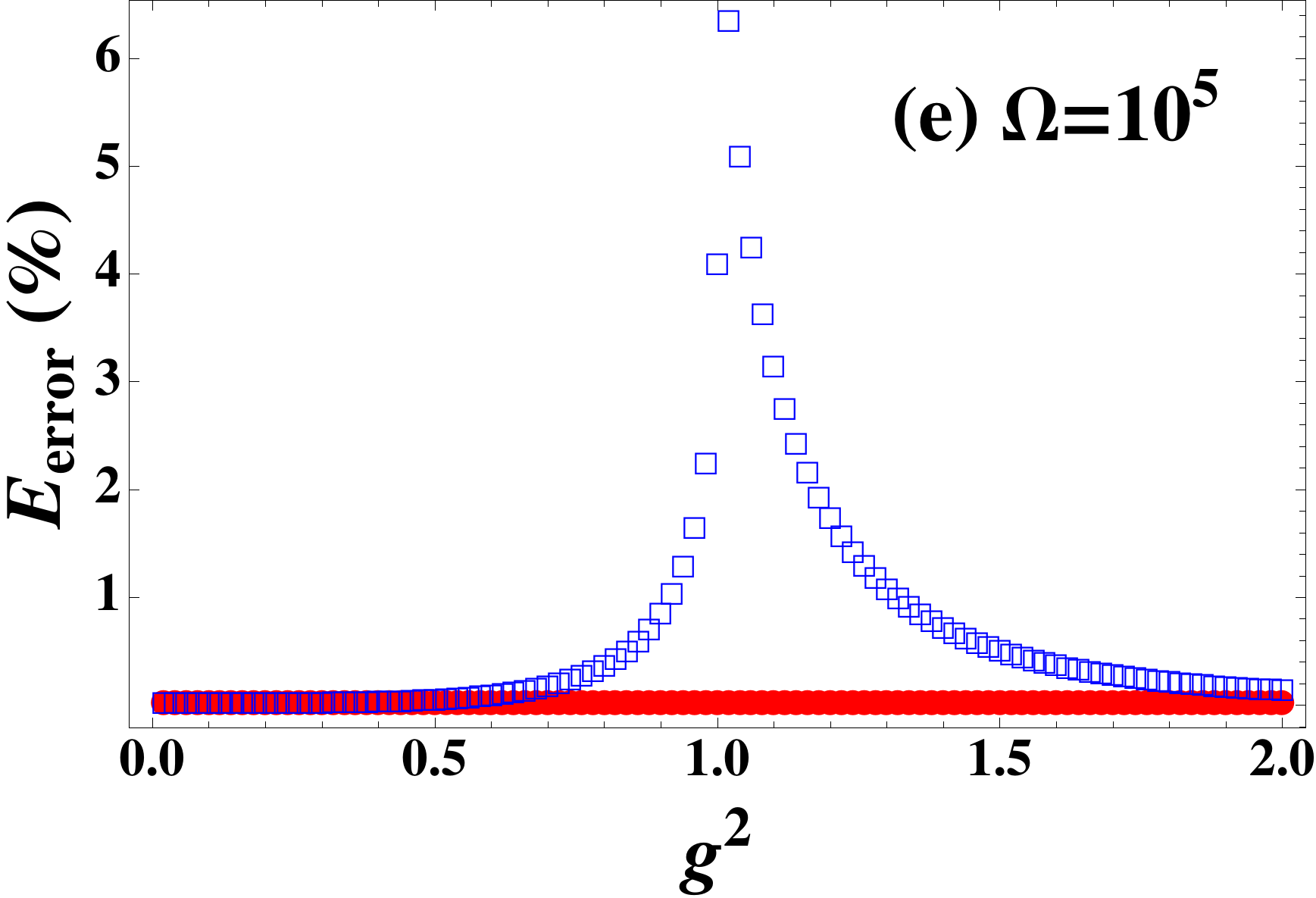} \\
\includegraphics*[width=42mm]{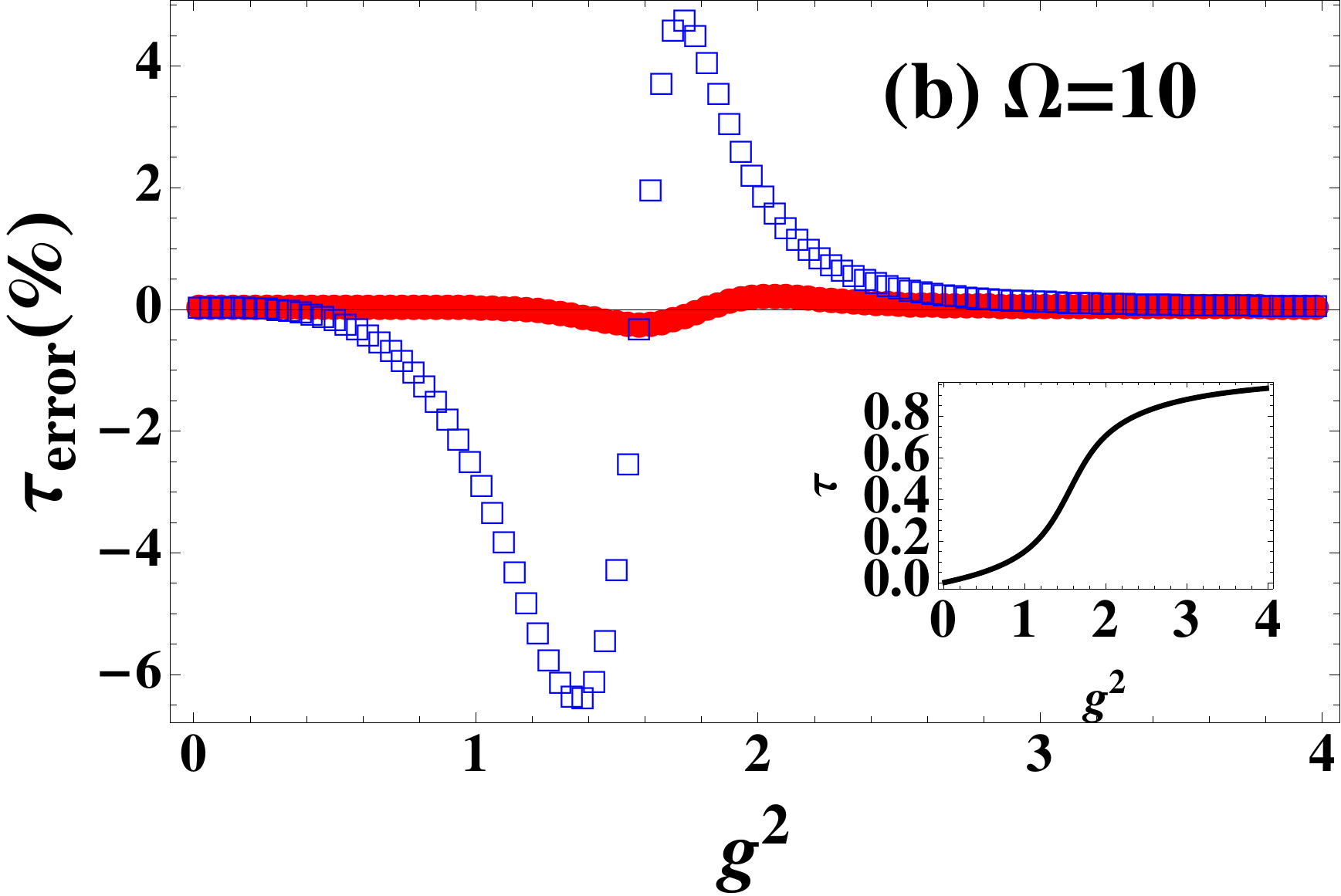} &
\includegraphics*[width=42mm]{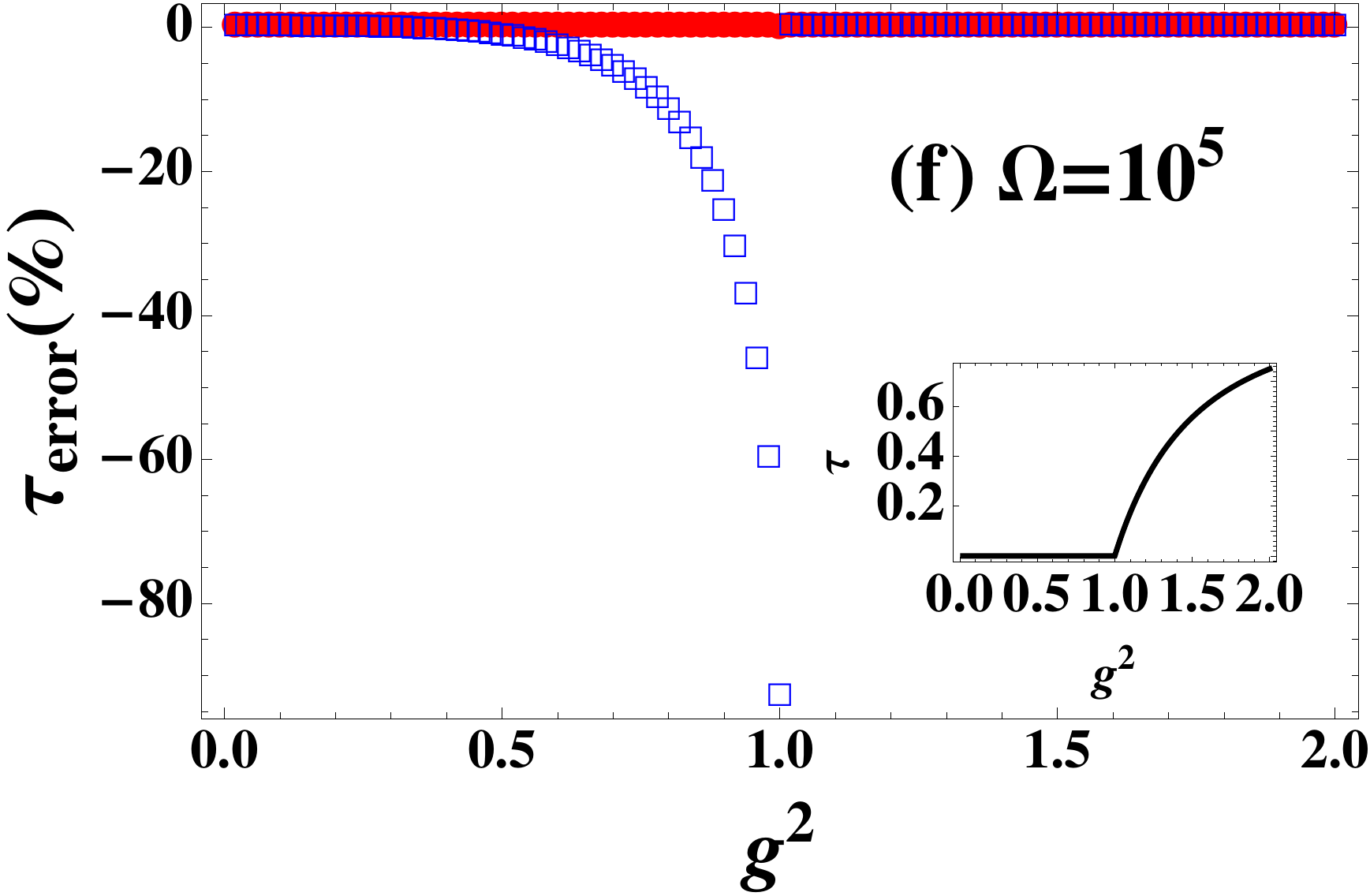} \\
\includegraphics*[width=42mm]{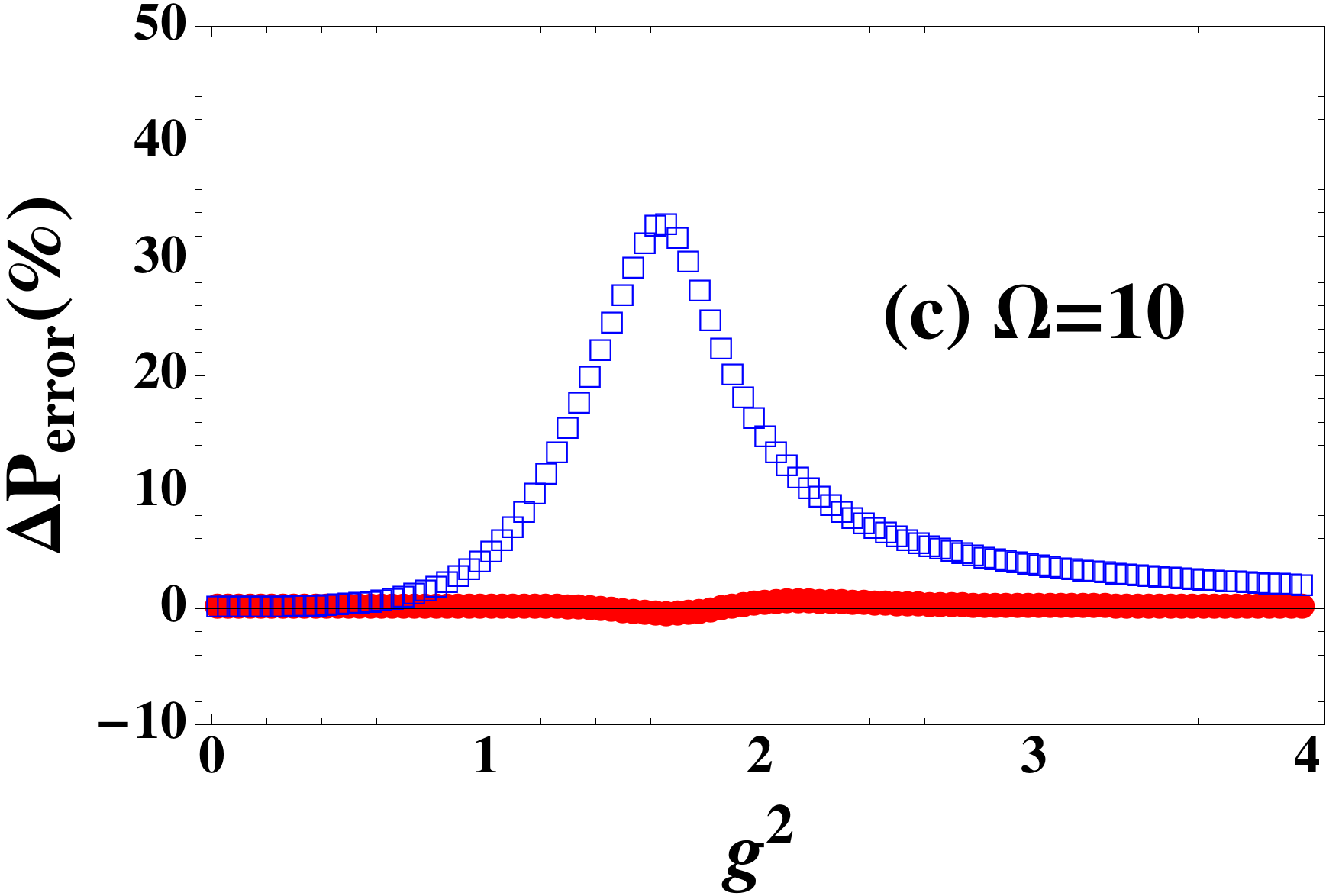} &
\includegraphics*[width=42mm]{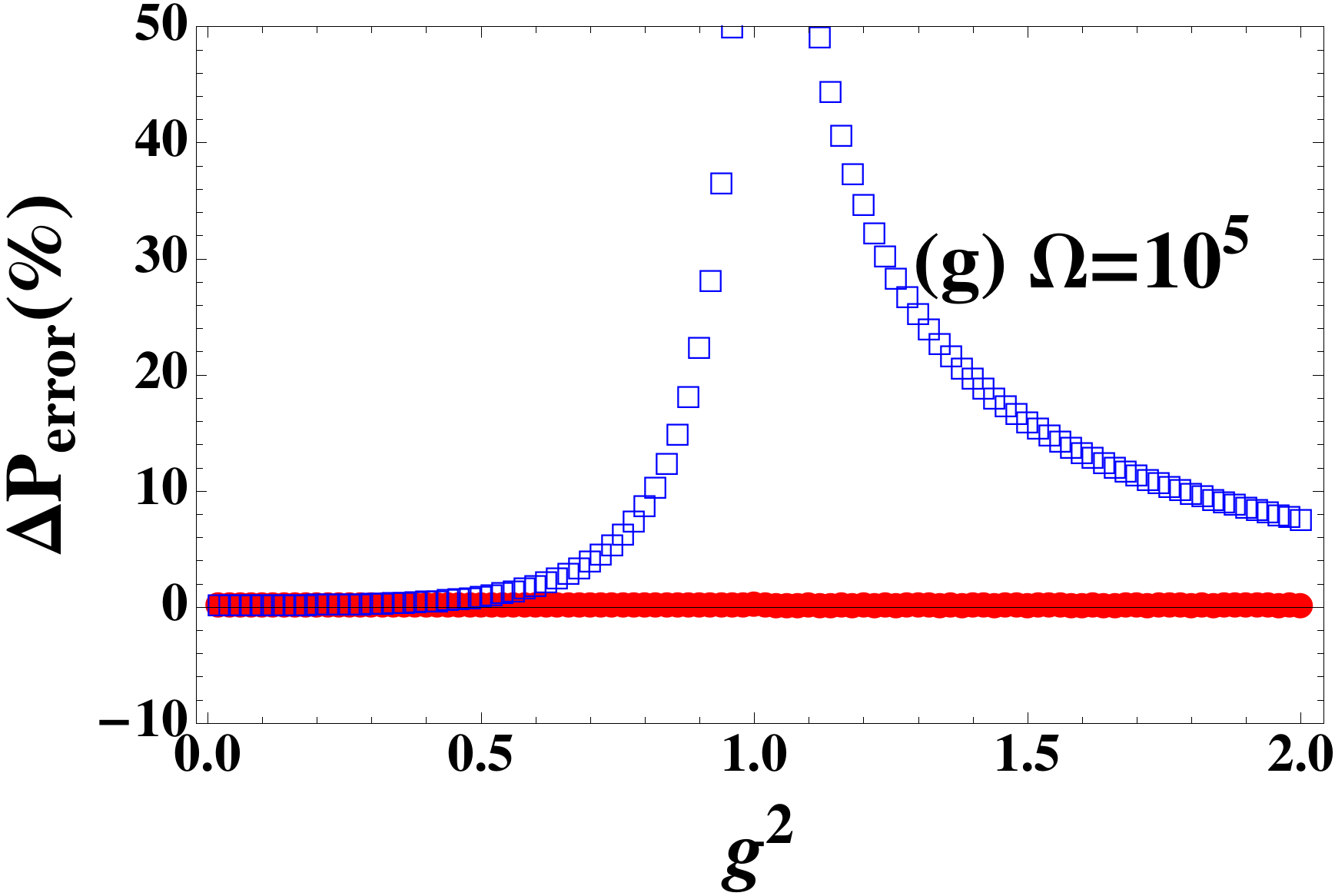} \\
\includegraphics*[width=42mm]{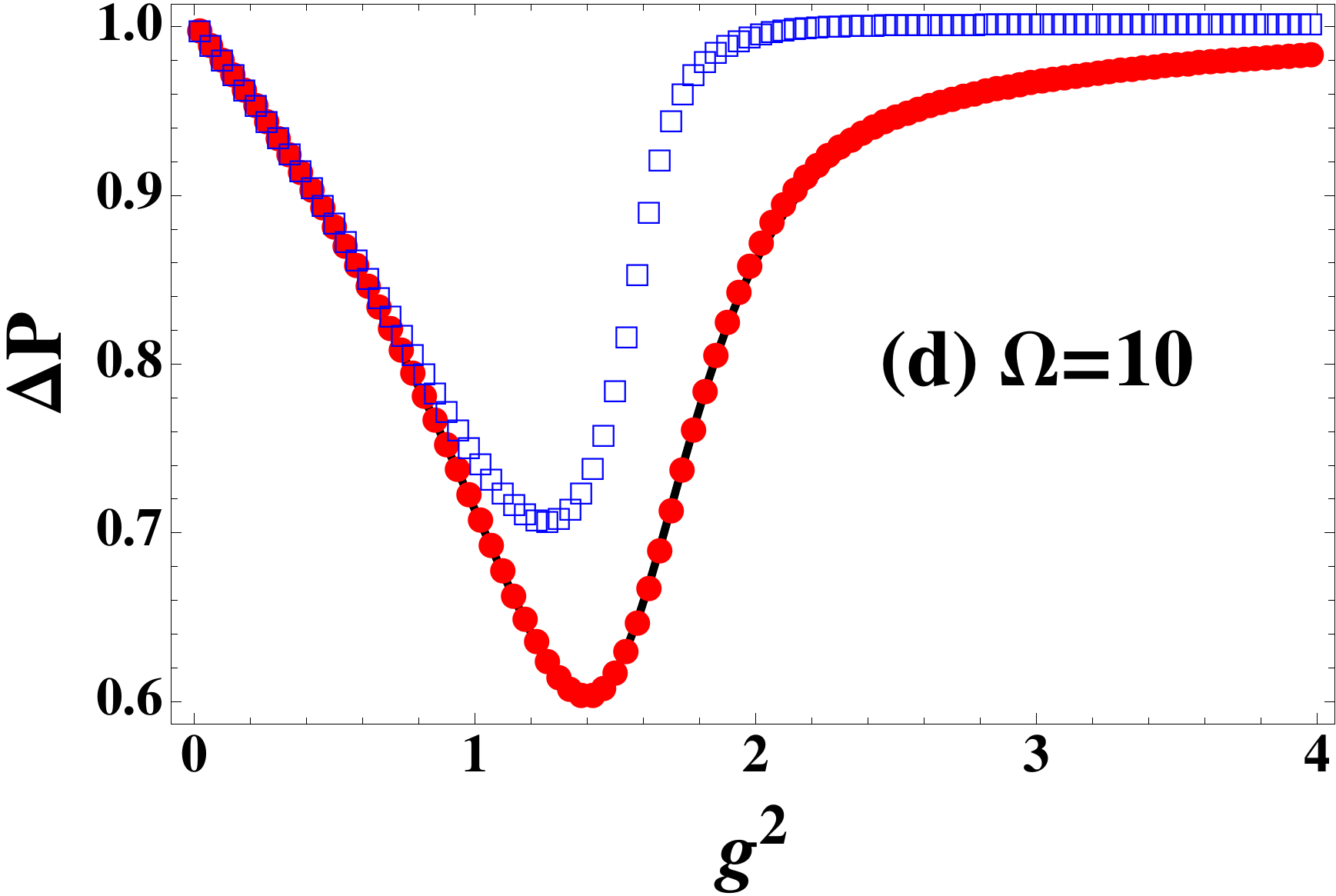} &
\includegraphics*[width=42mm]{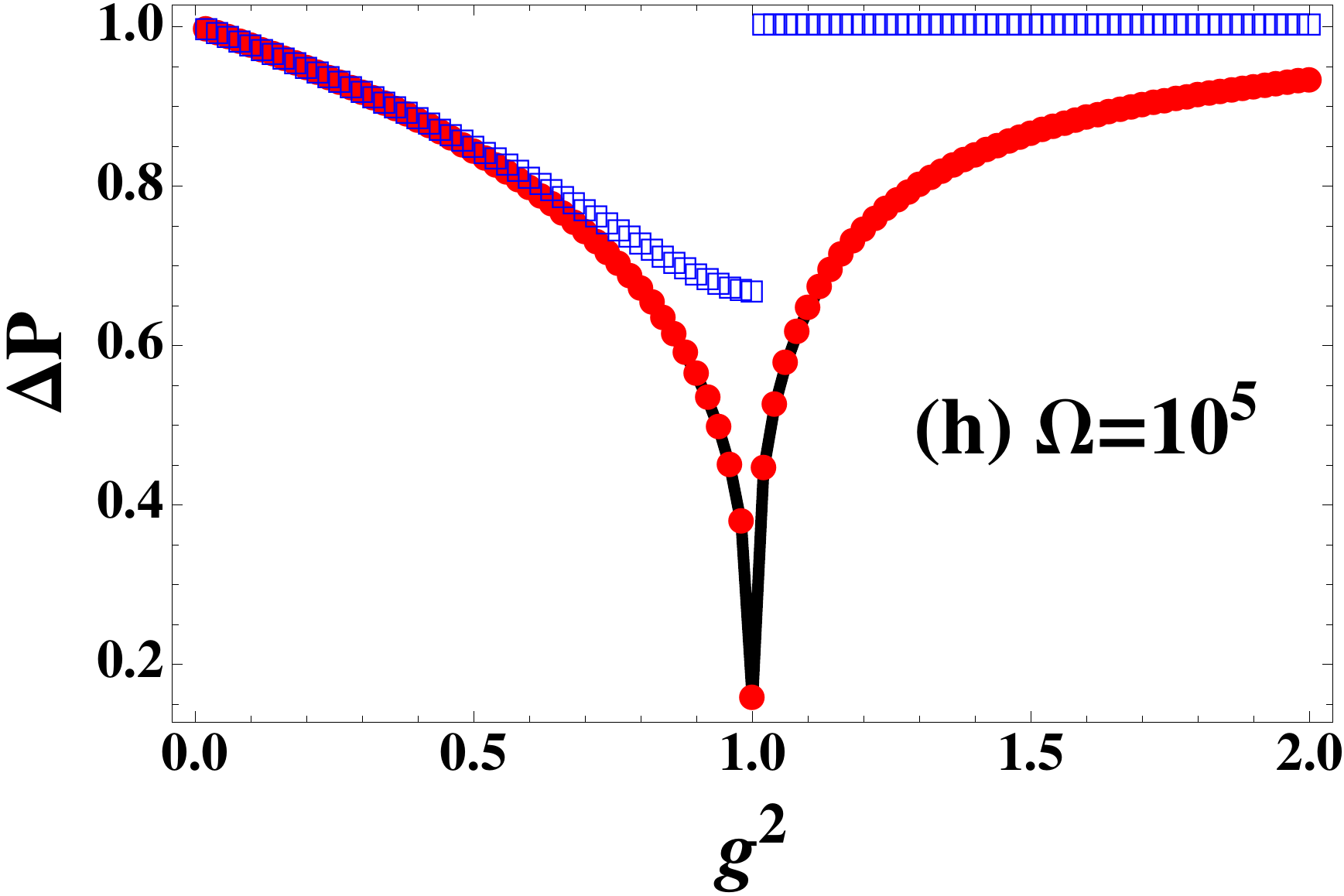}
\end{tabular}
\caption{(color online) Comparison of variational solutions with the exact
  solution. The (blue) empty square represents the DCS wave function, while
  the (red) filled circle indicates the DSS wave function. (a,e) Error in the
  energy with respect to the exact energy. (b,f) Error in the tangle. (c,g)
  Error in the momentum variance.  (d,h) The momentum variance of the ground
  state, the (black) line indicates the exact solution. }
\label{fig:2}
\end{figure}

In this paper, we develop a variational ground state. The variational method
had already been used for the Hamiltonian (\ref{Paper::eq:1}).  A displaced
coherent state for $g\gg 1$, a displaced squeezed state for $g\ll
1$~\cite{Chen89a}, and a superposition of two coherent states for
$g\simeq 1$~\cite{Shore73a,Stolze90a} have been suggested as a
variational ground state, respectively.  The accuracy of the suggested
variational states were reasonable on the ground-state
\emph{energy}~\cite{Chen89a,Shore73a,Stolze90a}.
However, we find that the variational states suggested in the previous
works~\cite{Shore73a,Stolze90a} underestimates the squeezing effect
significantly in the intermediate regime ($g\simeq1$); see
Figs.~\ref{fig:2} (b) and (e). The failure gets more pronounced as $\Omega$
increases.
% ; in the extreme limit of $\Omega\to\infty$, the squeezing effect in
% the variational ground state suddenly vanishes at $g\geq1$ whereas the
% exact solution shows rather a smooth change.

This observation implies that \emph{an energy-optimized variational
  wave function does not necessarily capture all the quantum correlation
  effects} in the true ground state of the system.
% In the particular regime
% ($g\simeq 1$) of interest, a good variational wavefunction must capture
% the squeezing effect accurately.
%%
In this paper, we suggest a new variational ground state which captures
accurately the squeezing effect.
% As one of the important physical quantity to be
% measured in the ultrastrong regime is the momentum variance, it is important
% to have a better approximate solution for the ground state, which is
% established in this paper.

First of all, we derive an equivalent model with only a boson degrees of
freedom, followed by a derivation of an effective classical Hamiltonian by
utilizing a parity symmetry in the model~\cite{Benivegna87a}. We observe that
this effective classical Hamiltonian shows bifurcation around $g=1$,
which is consistent with the observation in Ref.~(\cite{Hines04a,Meaney10a})
based on a classical model of the Eq.~(\ref{Paper::eq:1}), and
gives us qualitative account for the properties of the ground state.
We then suggest a superposition of two displaced-squeezed state as a trial
wave function, and demonstrate that it can predict the squeezing effect
accurately.  This shows that a deformation of each superposed wave packet is
crucial to the squeezing effect of ground state. We also discuss the drastic
change in the entanglement degree as $g$ varies across $g=1$ when
$\Omega\rightarrow\infty$.

\section{Effective Hamiltonian}

% The Hilbert space %of the Hamiltonian (\ref{Paper::eq:1})
% is spanned by tensor products of the boson number state $\ket{n}$
% ($n=0,1,2,\cdots$), $a^\dag{a}\ket{n}=n\ket{n}$, and the spin state
% $\ket\sigma$ ($\sigma=\pm1$), $\sigma_z\ket{\sigma}=\sigma\ket{\sigma}$.
% The
% zero energy is set to be $\lambda^2$ for the sake of a computational
% convenience.

The model (\ref{Paper::eq:1}) has a useful symmetry.  If
$\ket{n',\sigma'}=H\ket{n,\sigma}$, then
$n'+\sigma'\pmod{2}=n+\sigma\pmod{2}$, where
$\ket{n,\sigma}\equiv\ket{n}\otimes\ket{\sigma}$, $\ket{n}$ ($n=0,1,2,\cdots$)
is the boson number state, $a^\dag{a}\ket{n}=n\ket{n}$, and $\ket{\sigma}$
($\sigma=\pm1$) the spin state, $\sigma_z\ket{\sigma}=\sigma\ket{\sigma}$.  We
further stress this symmetry by introducing a generalized ``parity'' operator
\begin{equation}
\label{Paper::eq:2}
\Pi = \exp(-i\pi a^\dag a)\sigma_z \,.
\end{equation}
Clearly, the Hamiltonian~(\ref{Paper::eq:1}) is invariant under transformation
described by $\Pi$, $[H,\Pi]=0$.  The operator $\Pi$ has two eigenvalues
$\pm1$, and corresponding eigenstates are given by
\begin{equation}
\label{Paper::eq:3}
\ket{\varphi_n^\sigma} \equiv
\frac{(a^\dag)^n\sigma_x^n}{\sqrt{n!}}\ket{0,\sigma} \quad
(\sigma=\pm 1)
\end{equation}
Namely,
\begin{math}
\Pi\ket{\varphi_{n}^\pm} = \pm\ket{\varphi_n^\pm}
\end{math}.
The Hilbert space can be decomposed into a direct sum
$\calE=\calE_+\oplus\calE_-$ of two subspaces $\calE_\pm$ spanned by
$\ket{\varphi_n^\pm}$, respectively.  Accordingly, the Hamiltonian can also be
written as $H=H_+\oplus H_-$, where $H_\pm$ belongs to the subspace
$\calE_\pm$, and the partition function $Z\equiv\Tr[e^{-\beta{H}}]$ as
$Z=Z_++Z_-$, where $Z_\pm\equiv\Tr[e^{-\beta{H}_\pm}]$.

Now we make a key observation, suggested in the expression for the
parity eigenstates in Eq.~(\ref{Paper::eq:3}), that the combination
$a^\dag\sigma_x$ behaves like a boson operator within each subspace
$\calE_\pm$.  To show this rigorously, we first take a drone-fermion
representation~\cite{Spencer67a} of spin
\begin{math}
\sigma_z = 2f^\dag f - 1
\end{math}
and
\begin{math}
\sigma_+= \sigma_-^\dag = f^\dag(d+d^\dag),
\end{math}
where $f$ and $d$ are fermion operators. We then define a new boson operator
\begin{math}
b^\dag = a^\dag
(f^\dag-f)(d + d^\dag)
\end{math}
which satisfies $[b,b^\dagger]=1$.  The parity basis states are then written
as
\begin{equation}
\ket{\varphi_n^\sigma} = \frac{(b^\dag)^n}{\sqrt{n!}}\ket{0,\sigma} \,,
\end{equation}
where $\ket{0,\sigma}$ serves as the ``vacuum'' of the new boson operator $b$
within $\calE_\sigma$.  It is now clear that the sub-block Hamiltonians
$H_\pm$ can be rewritten solely in terms of the boson operator $b$ as
\begin{equation}
\label{Paper::eq:4} H_{\pm} = (b^\dag-\lambda)(b-\lambda) \pm
\frac{1}{2}\Omega \cos(\pi b^\dag b) \,.
\end{equation}
The same expression was derived also in Ref.~\cite{Benivegna87a} using a
canonical transformation. Having only a boson field, the
expression~(\ref{Paper::eq:4}) allows us to compare the model directly with
corresponding classical system by deriving the classical effective potential.
\begin{figure}
\centering
\includegraphics*[width=40mm]{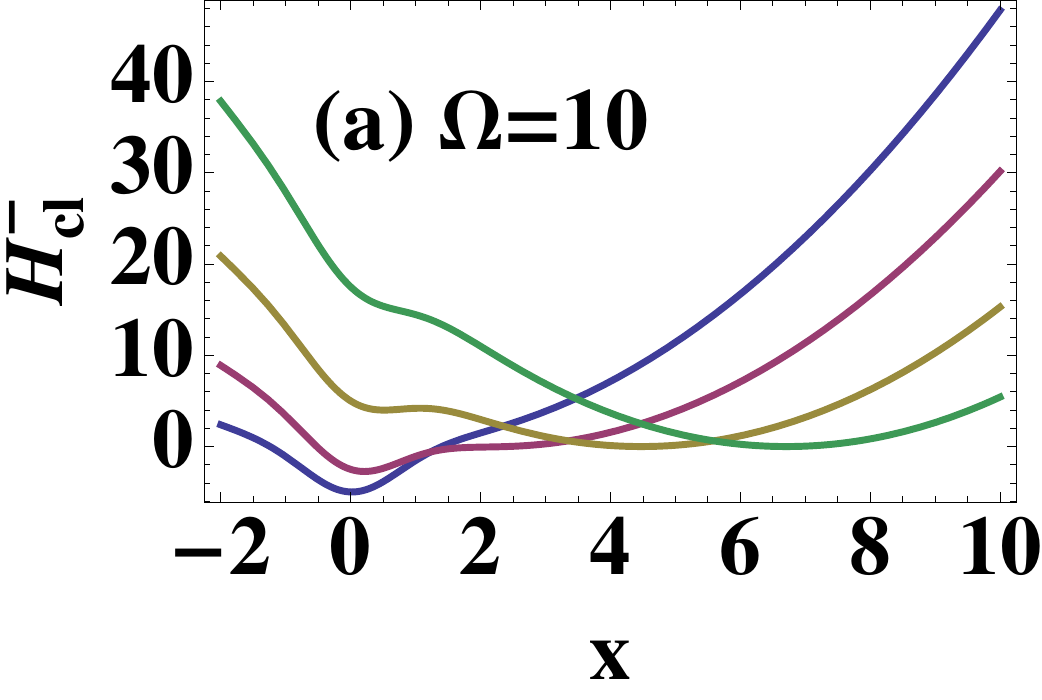}
\includegraphics*[width=40mm]{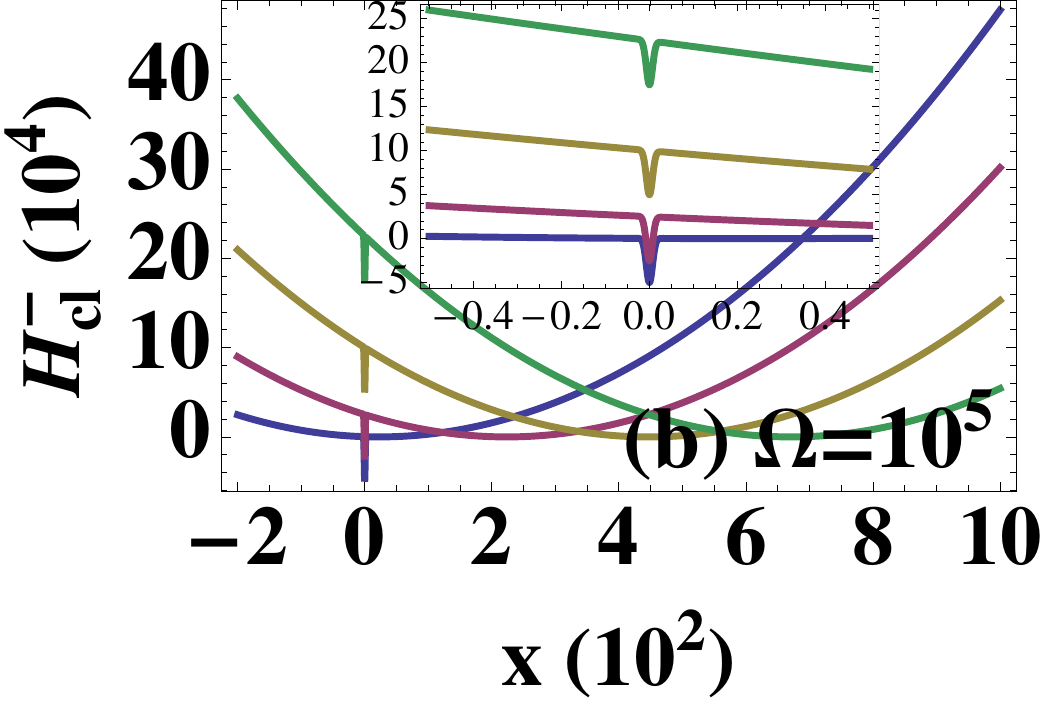}
\caption{(color online) The profile of $H_\cl^{-}(x,0)$ (a) for $\Omega=10$
  and (b) for $\Omega=10^5$. Different curves correspond to $g=0.1,1,2,3$
  (from bottom to top).  The inset in (b) zooms in the region around $x=0$.}
\label{fig:1}
\end{figure}

To get a classical effective potential, we express the partition
functions $Z_\pm$ in terms of a functional
integration~\cite{Negele88a}.
\begin{math}
Z_\pm=\int\calD[\phi^*(\tau),\phi(\tau)]\exp(-S_\pm)
\end{math}
with the action defined by
\begin{equation}
S_\pm = \int_0^\beta{d\tau}
\left[\phi^*\partial_\tau\phi+(\phi^*-\lambda)(\phi-\lambda)
  \pm \frac{1}{2} \Omega e^{-2\phi^*\phi}
\right]
\end{equation}
where $\beta$ is the inverse temperature, $\partial_\tau$ denotes
the partial derivative with respect to the imaginary-time $\tau$.
Making a change of variables, $p=(\phi^*+\phi)/\sqrt{2}$ and
$p=i(\phi^*-\phi)/\sqrt{2}$, we can rewrite the functional integrals
as
\begin{math}
Z_\pm = \int\mathfrak{\calD}[x(\tau),p(\tau)]\exp(-S_\pm)
\end{math}
with
\begin{eqnarray}
S_\pm = \int_0^\beta{d\tau} \left[-ip\partial_\tau x +
H_\cl^\pm(x,p)\right].
\end{eqnarray}
Here $H_\cl^\pm(x,p)$ is the effective classical Hamiltonian defined
by
\begin{equation}
\label{Paper::eq:5} H_\cl^\pm(x,p) =
\frac{1}{2}\left[(x-\sqrt{2}\lambda)^2+p^2\right] \pm
\frac{1}{2}\Omega e^{-(x^2+p^2)}
\end{equation}

The potential profile is depicted in Fig.~\ref{fig:1} for both $\Omega=10$ and
$\Omega=10^5$. For $g\ll 1$, the classical potential has one local minimum and
is highly anharmonic illustrating why a squeezed state localized around $x=0$
is a good variational solution in this regime. As $g$ approaches the critical
value, 1, it shows a bifurcation of its local minimum as in the corresponding
classical system.  As $g$ increases further, its local minimum around the
origin disappears and a single local minimum develops at
\begin{math}
x\sim g\sqrt{\Omega/2}.
\end{math}
It is also consistent with the fact that the coherent
state is a good approximation in this regime.

\section{Variational Solutions}

Above we have seen from the effective classical potential that the model in
Eq.~(\ref{Paper::eq:1}) shows highly non-trivial behaviors in the regime
$g\sim1$.  Now we investigate more closely this regime in terms of variational
wave functions.  To test the accuracy of a given trial wave function, we will
examine not only the energy but also the quantum correlation effects such as
squeezing and spin-oscillator entanglement.  We characterize the squeezing
effect with the momentum variance $\Delta{P}$ and the spin-oscillator
entanglement with \emph{tangle}~\cite{Coffman00a} defined by
\begin{math}
\tau=2[1-\Tr(\rho_s^2)],
\end{math}
where $\rho_s$ is a reduced density matrix of the two-level system.
We test the variational solution for two cases $\Omega=10$ and $\Omega=10^5$
by comparing it with the exact solution.
The exact solution is numerically calculated by the exact diagonalization
method \cite{endnote1}, where a sufficient number of Fock basis states
$\ket{n}$ (typically 300 states for $\Omega=10$ and 3000 states for
$\Omega=10^5$) are kept until the desired accuracy is achieved.

In Refs.~\cite{Chen89a,Stolze90a}, a superposition of two displaced coherent
states was suggested as a trial wave function,
\begin{equation}
\label{Paper::eq:5}
\ket{\psi(\alpha_1,\alpha_2,t)} =
(1-t)|\alpha_1\rangle + t|\alpha_2\rangle,
\end{equation}
where
\begin{math}
\ket{\alpha}=D(\alpha)\ket{0}
\end{math}
with $D(\alpha) = \exp[\alpha (b^\dag-b)]$.  Hereafter we will refer to the
trial wave function in Eq.~(\ref{Paper::eq:5}) as the \emph{double coherent
  state} (DCS).  It shows a reasonable accuracy for the energy (with error
about 1\%) in the intermediate regime [see Figs.~\ref{fig:2} (a) and (e)].
Although neither our effective classical potential nor a bifurcation of fixed
point in the corresponding classical system can give an quantitative account
for the ground state, the fact that a superposition of two wave packet is a
good approximation for the ground state is consistent with the classical
observations of bifurcation.

However, as shown in Figs.~\ref{fig:2} (c) and (f), the degree of entanglement
between the spin and oscillator exhibits relatively larger deviations,
especially for large $\Omega$.

The poor accuracy of the variational solution in Eq.~(\ref{Paper::eq:5}) is
even more noticeable (more than 30\%) on the momentum variance $\Delta{P}$ of
the ground state. For both $\Omega=10$ and $10^5$ case, the squeezing effect
is considerably underestimated in this scheme. Moreover, for the true
adiabatic regime, the variational solution predicts that the squeezing effect
suddenly disappears when $g$ becomes larger than 1.

\begin{figure}
\begin{tabular}{cc}
\includegraphics*[width=42mm]{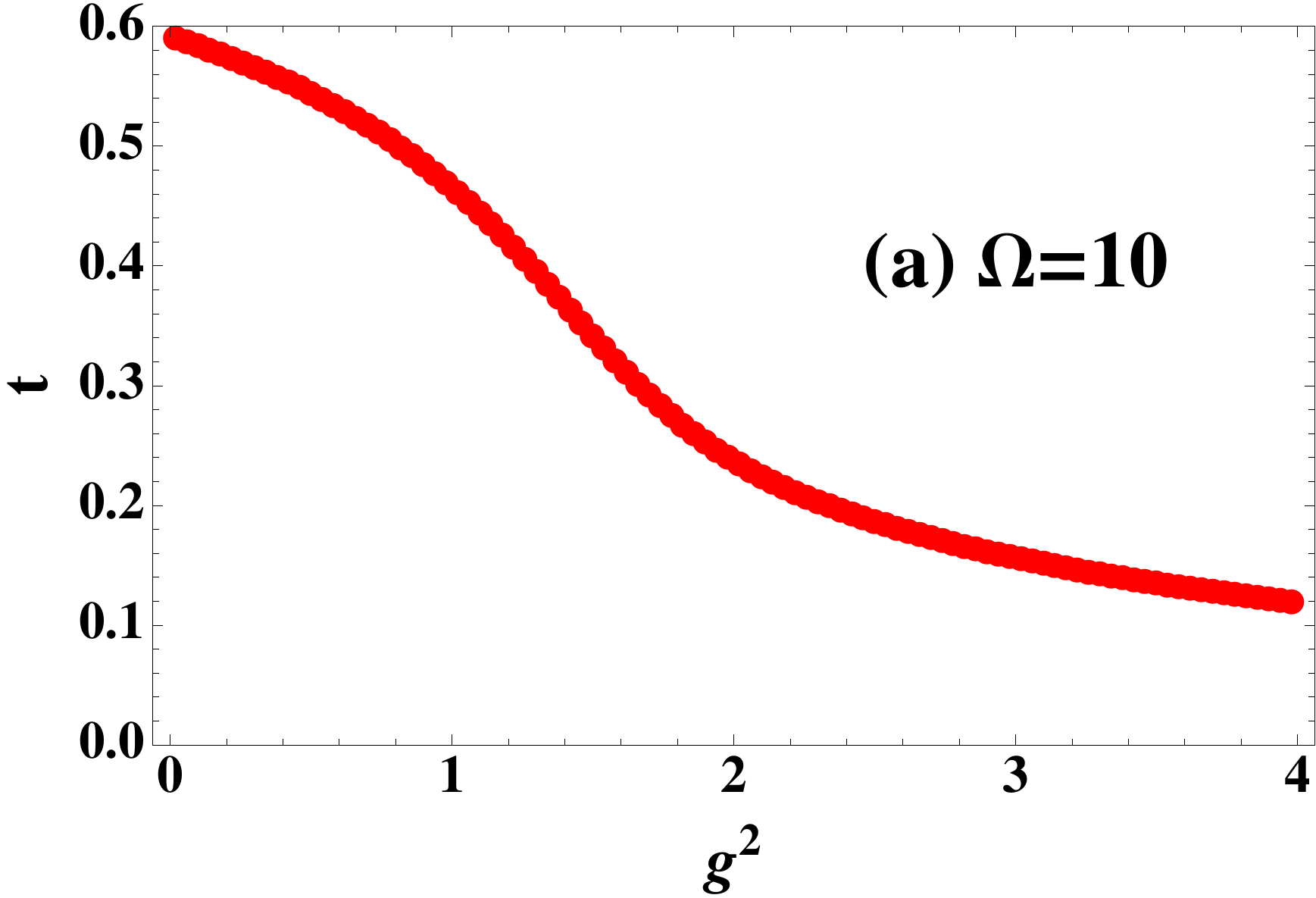} &
\includegraphics*[width=42mm]{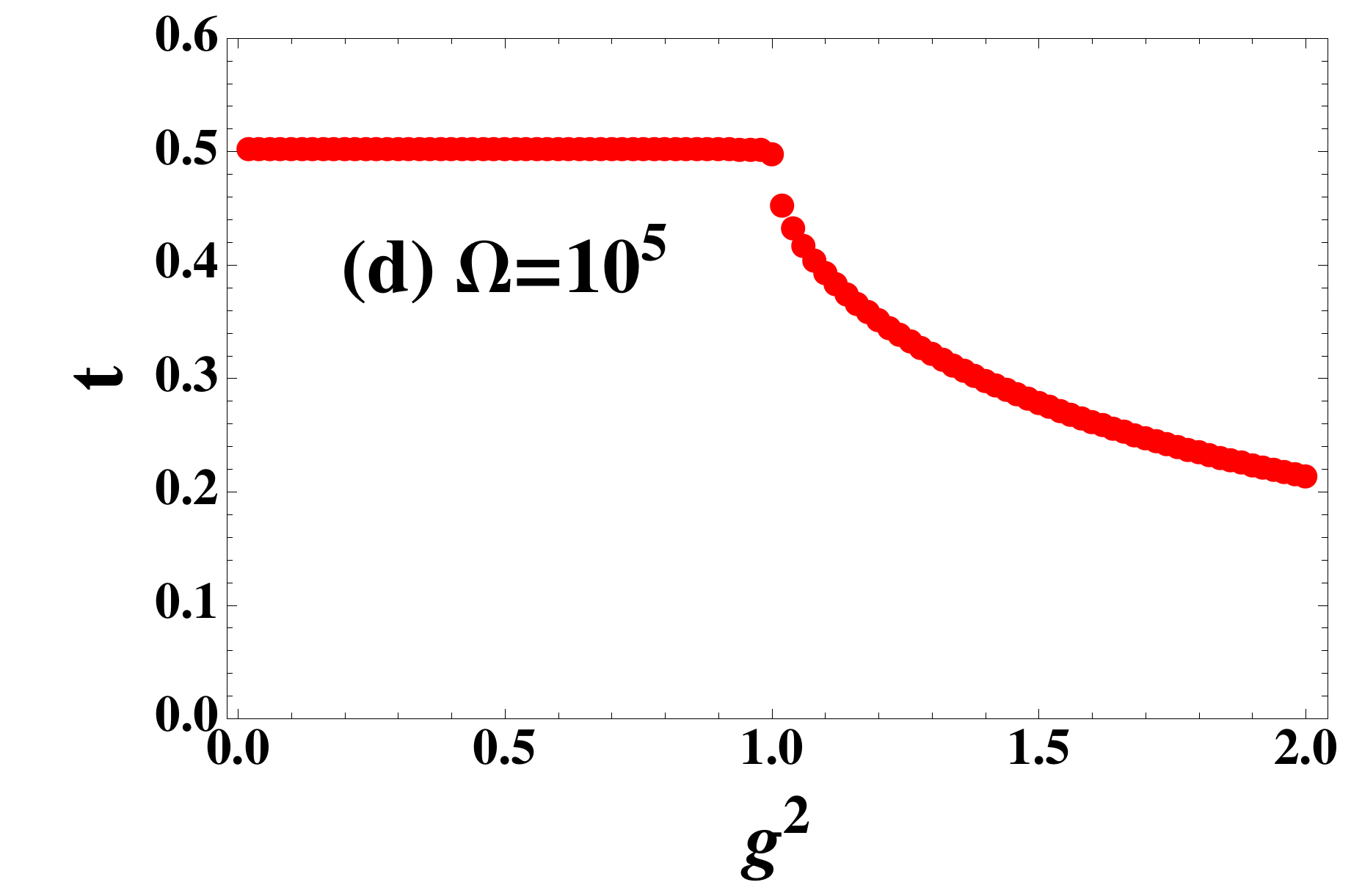} \\
\includegraphics*[width=42mm]{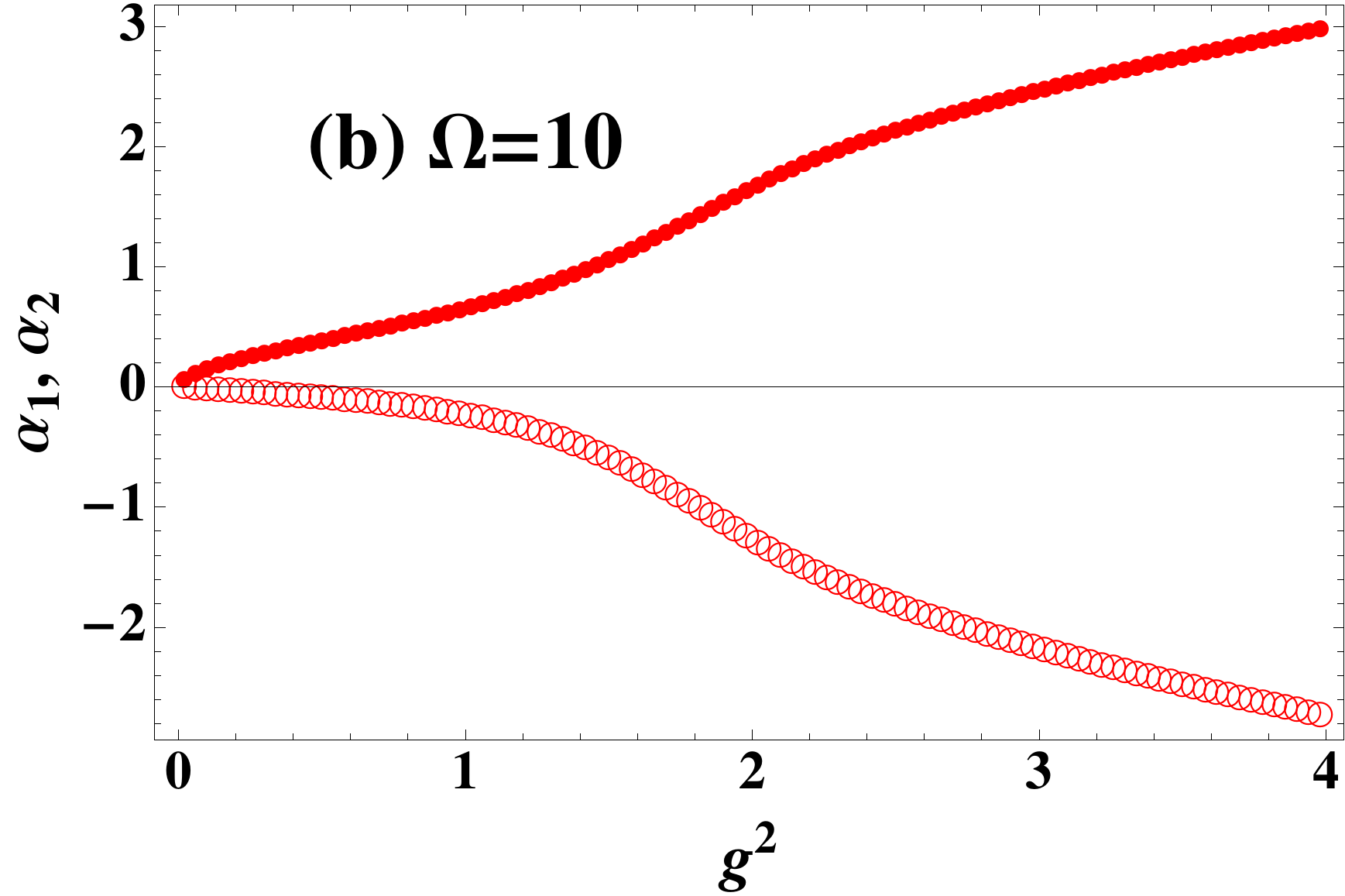} &
\includegraphics*[width=42mm]{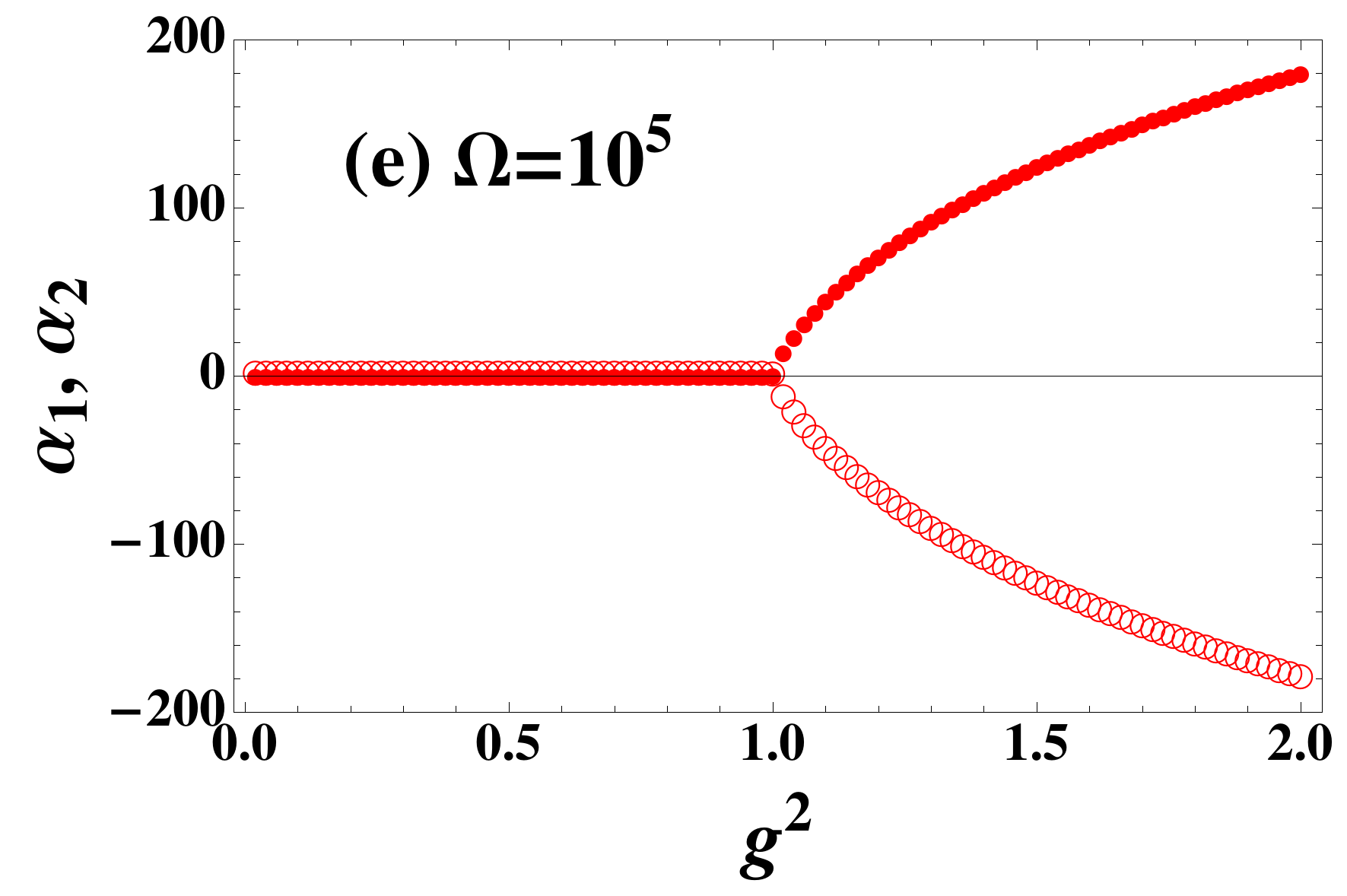} \\
\includegraphics*[width=42mm]{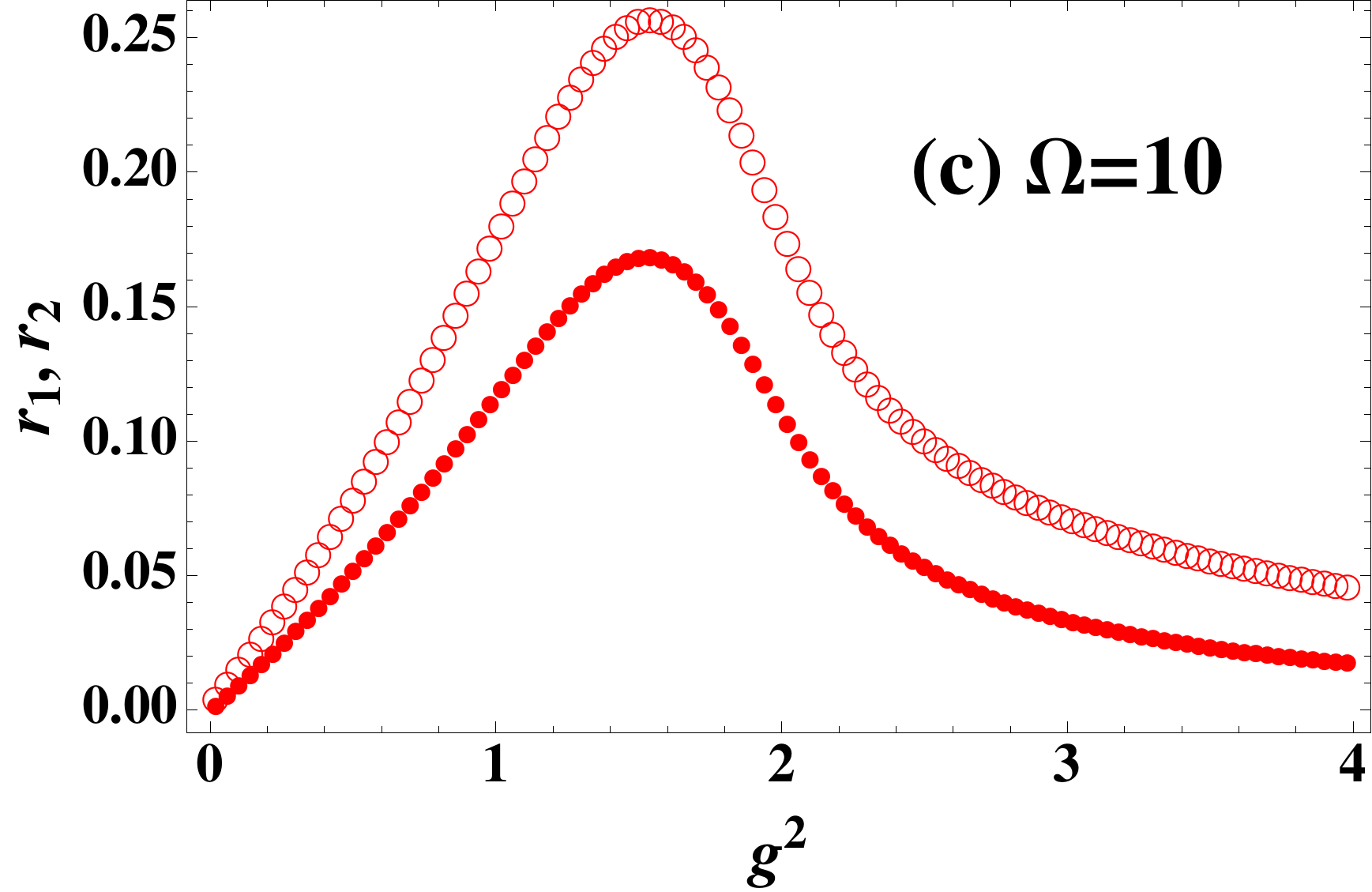} &
\includegraphics*[width=42mm]{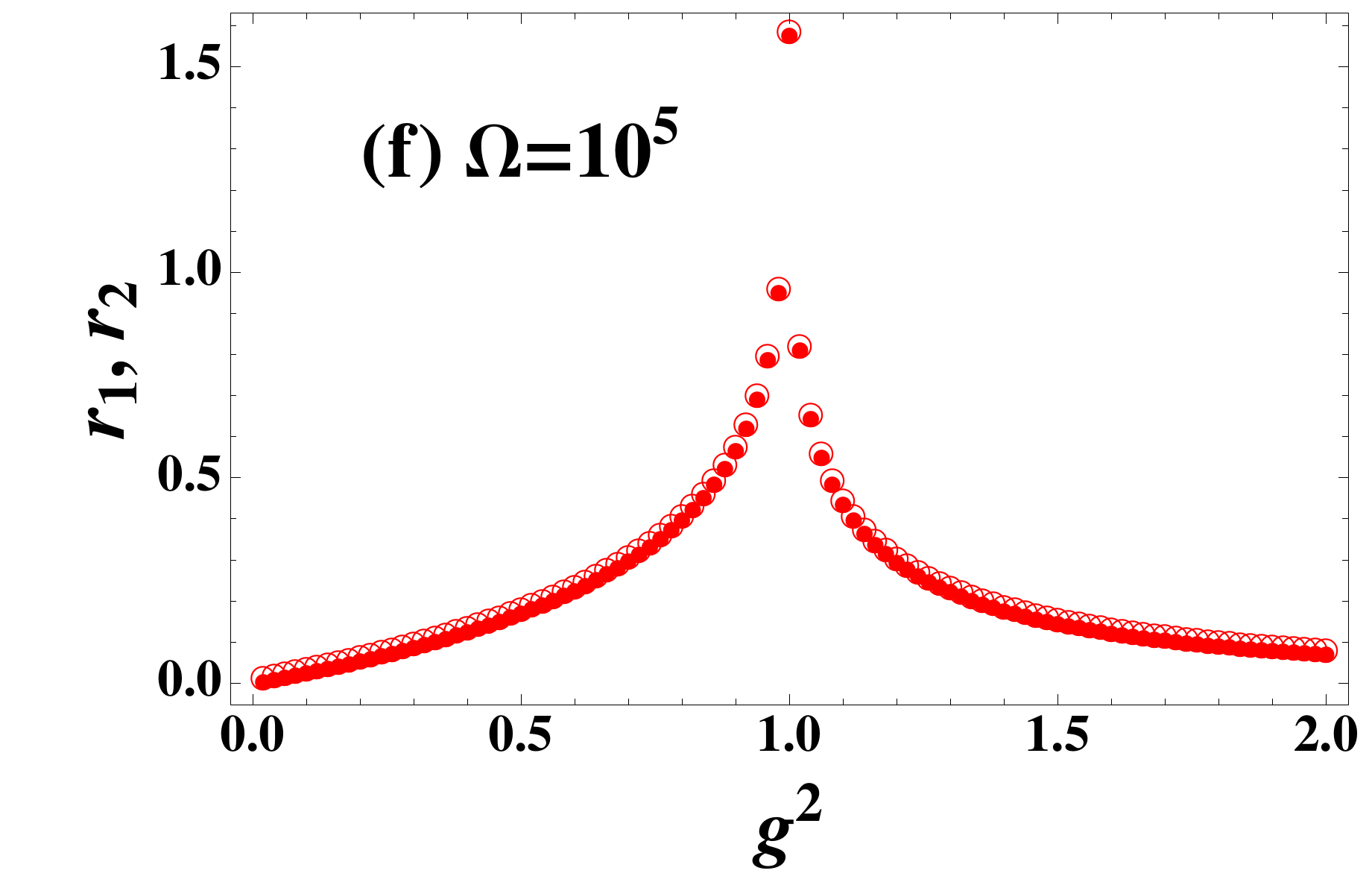} \\
\end{tabular}
\caption{(color online) Optimized variational parameters for the \emph{double
    squeezed state} [Eq.~(\ref{Paper::eq:5})] for $\Omega=10$ (a,b,c) and for
  $\Omega=10^5$ (d,e,f).  The variational parameters for the \emph{double
    coherent state} [Eq.~(\ref{Paper::eq:6})] are qualitatively similar (not
  show) except for $r_1=r_2=0$. (a,d) The relative weight $t$ in the
  superposition. (b,e) The displacement parameters $\alpha_1$ and
  $\alpha_2$. (c,f) The squeezing parameters $r_1$ and $r_2$.}
\label{fig:3}
\end{figure}

We can understand the reason for the failure by looking at the optimized
variational parameters shown in Fig.~\ref{fig:3} (e). If $g\ll1$, the
variational states are optimized at
\begin{math}
\alpha_1=-\alpha_2\sim g\sqrt{\Omega}/2.
\end{math}
In this case, there is a
considerable overlap between two coherent states which results in the
squeezing effect.  On the contrary, when $g$ becomes larger than 1, the
variational parameter abruptly changes, and the value for $\alpha_1$ and
$\alpha_2$ becomes an order of 10 or 100 still having opposite signs with each
other. Then, the overlap between two coherent states vanishes, and cannot give
the squeezing effect. This observation lead us to conclude that a
deformation of individual wave packets is considerable in the true ground
state, and unless the effect is taken into account in the trial wave function,
the squeezing effect is inevitably underestimated.

Motivated from the above observation, we propose a superposition of
two displaced-squeezed states as a trial wave function,
\begin{equation}
\label{Paper::eq:6}
\ket{\psi(r_1,\alpha_1,r_2,\alpha_2,t)} =
(1-t)|r_1,\alpha_1\rangle + t|r_2,\alpha_2\rangle,
\end{equation}
where $\ket{r,\alpha}$ denotes the displaced-squeezed
state~\cite{Scully97a} defined by
$\ket{r,\alpha}=S(r)\ket{\alpha}$ with
\begin{math}
S(r)=\exp[(rb^\dagger b^\dagger-r^*bb)/2] .
\end{math}
We refer to the wave function in Eq.~(\ref{Paper::eq:6}) as the \emph{double
  squeezed state} (DSS).
Without a loss of generality, the displacement parameter $\alpha$ and the
squeezing parameter $r$ are assumed to be real.

The variational parameters $r_1,\alpha_1,r_2,\alpha_2$, and $t$ are
determined by minimizing the energy
\begin{math}
E(r_1,\alpha_1,r_2,\alpha_2,t) =
\langle\psi|H_-|\psi\rangle/\langle\psi|\psi\rangle
\end{math}.
It includes two direct terms,
\begin{equation}
\langle r_j,\alpha_j|H_-|r_j,\alpha_j\rangle=\sinh^2r_j
+ (\lambda-\alpha_je^{r_j})^2
- \frac{\Omega}{2}e^{-2\alpha_j^2}
\end{equation}
for $j=1,2$,  and a cross term
\begin{multline}
\frac{\braket{r_1,\alpha_1|H_1|r_2,\alpha_2}}{
  \braket{r_1,\alpha_1|r_2,\alpha_2}}
= \sech(r_1-r_2)\sinh{r_1}\sinh{r_2} \\{} +
\left[\lambda-\sech(r_1-r_2)(\alpha_1\cosh{r_2}+\alpha_2\sinh{r_1})\right]
\hfil\\{}\times
\left[\lambda-\sech(r_1-r_2)(\alpha_1\sinh{r_2}+\alpha_2\cosh{r_1})\right]
\\{}
- \half\Omega\exp\left[-2\sech(r_1-r_2)\alpha_1\alpha_2\right]
\end{multline}
where
\begin{multline}
\braket{r_1,\alpha_1|r_2,\alpha_2}
= \frac{\exp\left[\alpha_1\alpha_2\sech(r_1-r_2)\right]}{
  \sqrt{\cosh(r_1-r_2)}} \\{}\times
\exp\left[-\frac{(\alpha_1^2+\alpha_2^2)
    + \left(\alpha_1^2-\alpha_2^2\right)\tanh(r_1-r_2)}{2}\right]
\end{multline}
The optimum variational parameters $r_1$, $r_2$, $\alpha_1$, $\alpha_2$, and
$t$ thus determined are plotted as a function of $g$ in
Fig.~\ref{fig:3}.
The squeezing parameters $r_1$ and $r_2$ of the individual wave packets in the
DSS is indeed large in the regime $g\simeq1$, where the DCS variational
solution changes abruptly.
This is consistent with our prediction.

Using this variational solution, we calculate the momentum variance.  The
variance $\Delta{P}$ of the momentum $P=i(b^\dagger-b)$ is given by
\begin{math}
\Delta{P}=\braket{\psi|P^2|\psi}/\braket{\psi|\psi}
\end{math}
as in our case $\langle\psi|P|\psi\rangle=0$. The direct
terms and cross term are respectively given by
\begin{math}
\braket{r_j,\alpha_j|P^2|r_j,\alpha_j}=e^{-2r_j}
\end{math}
($j=1,2$)
and
\begin{equation}
\frac{\braket{r_1,\alpha_1|P^2|r_2,\alpha_2}}
{\braket{r_1,\alpha_1|r_2,\alpha_2}} = \frac{2}{e^{2r_1}+e^{2r_2}} -
\frac{4(\alpha_1e^{r_1}-\alpha_2e^{r_2})^2}{(e^{2r_1}+e^{2r_2})^2}.
\end{equation}
Thus calculated momentum variation is shown in Fig.~\ref{fig:2} (d) and
(h). The squeezing effect is accurately captured by the DSS
(\ref{Paper::eq:6}), and the accuracy can be attributed to the deformation of
the constituent wave packets.
Our variational solution illustrates well the phase transition-like
behavior for $\Omega\rightarrow\infty$ regime.
% For simplicity,
% let us consider the DCS wavefunction in Eq.~(\ref{Paper::eq:5}).
% If we express the state in the original Hilbert space composed with
% the boson and the spin, it is
% \begin{eqnarray*}
% |\psi\rangle
% &=&\sum_{n=\textrm{even}}^\infty\frac{\alpha^n}{\sqrt{n!}}
% |n\rangle\ket{\downarrow}
% +(1-2t)\sum_{n=\textrm{odd}}^\infty\frac{\alpha^n}{\sqrt{n!}}
% |n\rangle\ket{\uparrow}
% \end{eqnarray*}
% As it is shown in Fig.~(\ref{fig:3}), the variational solution takes
% $t=1/2$ for $\alpha<1$, which means that the two coherent state
% reduces to
% \begin{equation}
% |\psi\rangle
% =\sum_{n=\textrm{even}}^\infty
% \frac{\alpha^n}{\sqrt{n!}}|n\rangle\|\downarrow\rangle
% \end{equation}
% which is clearly a separable state and gives a zero tangle. As the $g$
% becomes larger than the critical value, $t$ deviates from $1/2$ and gives rise
% to sudden appearance of the entanglement.

Interestingly, the optimal variational parameters satisfy
$\alpha_1\approx-\alpha_2$ and $r_1\approx{}r_2$ for $\Omega\gg1$ [it is not
the case for $\Omega\sim1$; compare Fig.~\ref{fig:3} (b,c) with
Fig.~\ref{fig:3} (e,f)].  In this case, the DSS in Eq.~(\ref{Paper::eq:6}),
which has been expressed in terms of the spin-oscillator hybrid mode $b$,
takes a simple form (not normalized)
\begin{multline}
\ket{\psi(r,\alpha,r,-\alpha,t)}
= \left(\ket{r,\alpha}+\ket{r,-\alpha}\right)\otimes\ket{\downarrow} \\{}
+ (1-2t)\left(\ket{r,\alpha}-\ket{r,-\alpha}\right)\otimes\ket{\uparrow}
\end{multline}
in terms of the original spin and oscillator mode.  This form shows
transparently the entanglement feature in the optimal wave function: For
$g\ll 1$, $t\approx 1/2$ and the spin and oscillator is separable.  As $g$
increases, $t$ decreases, which leads to strong entanglement between spin and
oscillator mode as shown in the insets of Fig.~\ref{fig:2} (c) and (g).  and
the overlap between $\ket{\alpha}$ and $\ket{-\alpha}$ ($\alpha\sim\lambda$)
gives a finite amount of squeezing.

\section{Discussion}

We have studied the spin-boson model~(\ref{Paper::eq:1}) in the ultra-strong
coupling limit by means of an effective classical potential and an improved
variational wave function.

We note that in the true adiabatic regime ($\Omega\gg 1$), the critical
strength $g$ for the entanglement coincide with the point where the squeezing
effect is maximum. This coincidence can also be noticed from the crossover
behavior of the effective classical potential~(\ref{Paper::eq:5}).
Nevertheless, the precise relation among the bifurcation in the classical
potential in Fig.~(\ref{fig:1}) or in the corresponding classical
model~\cite{Hines04a}, the squeezing effect, and the entanglement remains an
interesting open problem in the ultra-strong coupling limit of the
model~(\ref{Paper::eq:1}).

% \begin{figure}
% \includegraphics*[width=6cm]{fig3a}
% \includegraphics*[width=6cm]{fig3b}
% \caption{A variational solution consisting of the superposition of two
%   \emph{coherent states} for $\Omega=10^5$. When $g$ becomes larger than
%   the critical value, 1, the absolute value $\alpha_1$ and $\alpha_2$ suddenly
%   increases. Then the overlap between the two wave packet decreases, which
%   leads to the vanishing squeezing effects shown in Fig.~\ref{fig:2} (e).  (b)
%   The relative weight $t$ in the superposition.  When $g<1$, $t$ is
%   always $1/2$, meaning that the ground state is separable.}
% \label{fig:3}
% \end{figure}

% \section{Conclusion}
% The superposition of the two displaced-squeezed states has been suggested as a
% variational wave function which can precisely predict the squeezing effect in
% the ultrastrong coupling regime. We demonstrated that the squeezing of the
% oscillator's momentum is accounted for by both the interference of the two
% wave packet and the deformation of the each wave packet. The variational
% solution in the true adiabatic regime shows what happens when the ground state
% undergoes the phase-transition like behavior in the entanglement.

% \begin{acknowledgments}
This work has been supported by the NRF Grant (No.~2009-0080453), the BK21
Program, and the APCTP.
% \end{acknowledgments}

\bibliographystyle{apsrev}
% \bibliography{alison,cience,physey,mathey,quaphy,conmat,staphy,mywork}
\bibliography{Paper}

\begin{thebibliography}{25}
\bibitem[{end()}]{endnote1}
\bibinfo{note}{One may also use recently proposed alternative methods in
  Refs.~\cite{Irish07a,1004.2801}.}
\end{thebibliography}

\end{document}